\documentclass[12pt]{article}
\usepackage{epsfig,here}
\usepackage{graphicx}
\usepackage{amssymb}
\usepackage{latexsym}
\usepackage{bbm}

\newcommand{\La}{\Lambda}
\newcommand{\eps}{\epsilon}

\newcommand{\nn}{\nonumber}
\newcommand{\nl}{\nonumber\\}

\newcommand{\dg}{\dagger}

\newcommand{\cL}{{\cal L}}
\newcommand{\cM}{{\cal M}}

\newcommand{\cH}{{\cal H}}
\newcommand{\gsim}{~{}_{\textstyle\sim}^{\textstyle >}~}
\newcommand{\lsim}{~{}_{\textstyle\sim}^{\textstyle <}~}
\newcommand{\ba}{\begin{array}{c}}
\newcommand{\bat}{\begin{array}{cc}}
\newcommand{\ea}{\end{array}}

\newcommand{\Dslash}{\mbox{${D\hspace{-.65em}/}$}}
\def\slch#1{\setbox0=\hbox{$#1$}\dimen0=\wd0%
\setbox1=\hbox{/}\dimen1=\wd1%
\ifdim\dimen0>\dimen1%
\rlap{\hbox to
\dimen0{\hfil/\hfil}}#1\else                                     
\rlap{\hbox to \dimen1{\hfil$#1$\hfil}}/\fi}
\newcommand{\lra}{\longrightarrow}
\begin{document}

\pagestyle{plain}

\parindent 0mm
\parskip 6pt

\vspace*{-1.8cm}
\begin{flushright}
UWThPh-2005-11 \\
July 2005
\end{flushright}

\vspace*{0.5cm} 
\begin{center}
{\Large\bf Effective Field Theories}\\[.2cm] 
Gerhard Ecker \\[.2cm] 
Institut f\"ur Theoretische Physik, 
Universit\"at Wien \\
Boltzmanng. 5, \,A-1090 Vienna, Austria\\
E-mail: gerhard.ecker@univie.ac.at
\end{center} 

\vspace*{1cm} 

\section{Introduction}

Effective Field Theories (EFTs) are the counterpart of the Theory of
Everything. They are the field theoretical implementation of 
the quantum ladder: heavy degrees of freedom need not be
included among the quantum fields of an EFT for a description of
low-energy phenomena. For example, we do not need 
quantum gravity to understand the hydrogen atom nor does chemistry
depend upon the structure of the electromagnetic interaction of
quarks. 

EFTs are approximations by their very nature. Once the relevant degrees
of freedom for the problem at hand have been established, the
corresponding EFT is usually treated perturbatively. It does not
make much sense to search for an exact solution of the Fermi theory of
weak interactions. In the same spirit, convergence of the perturbative
expansion in the mathematical sense is not an issue. The 
asymptotic nature of the expansion becomes apparent once the accuracy
is reached where effects of the underlying ``fundamental'' theory
cannot be neglected any longer. The range of applicability of the
perturbative expansion depends on the
separation of energy scales that define the EFT. 

EFTs pervade much of modern physics. The effective nature of the
description is evident in atomic and condensed matter
physics. The following article will be restricted to particle physics 
where EFTs have become important tools during the last 25 years. 

\section{Classification of effective field theories}
\label{class}

A first classification of EFTs is
based on the structure of the transition from the ``fundamental''
(energies $> \La$) to the ``effective'' level (energies $<
\La$).
\begin{enumerate} 
\item[1.] Complete decoupling.\\[.1cm]
The fundamental theory contains heavy and light degrees of freedom.
Under very general
conditions (decoupling theorem, Appelquist and Carazzone, 1975), the
effective Lagrangian for energies $\ll \La$, depending only on
light fields, takes the form
\begin{equation} 
\cL_{\rm eff} = \cL_{d\leq 4} + \sum_{d>4} \frac{1}{\La^{d-4}}
\sum_{i_d} g_{i_d} O_{i_d}~.  \label{L_EFT}
\end{equation}
The heavy fields with masses $> \Lambda$  have been ``integrated out'' 
completely. 
$\cL_{d\leq 4}$ contains the potentially renormalizable terms
with operator dimension $d \leq 4$ (in natural mass units where Bose
and Fermi fields have $d=1$ and 3/2, respectively), the $g_{i_d}$ are 
coupling constants and the $O_{i_d}$ are monomials
in the light fields with operator dimension $d$. In a slightly
misleading notation, $\cL_{d\leq 4}$ consists of relevant and 
marginal operators whereas the $O_{i_d}~(d>4)$ are denoted irrelevant 
operators. The scale $\La$ can
be the mass of a heavy field (e.g., $M_W$ in the Fermi theory
of weak interactions) or it reflects the short-distance structure 
in a more indirect way.
\item[2.] Partial decoupling.\\[.1cm]
In contrast to the previous case, the heavy fields do not disappear
completely from the EFT but only their high-momentum modes are
integrated out. The main area of application is the physics of heavy
quarks (Sec.~\ref{HQP}). The procedure involves one or several 
field redefinitions introducing a frame
dependence. Lorentz invariance is not manifest but implies relations
between coupling constants of the EFT (reparametrization invariance).  
\item[3.] Spontaneous symmetry breaking.\\[.1cm]
The transition from the fundamental to the effective level occurs
via a phase transition due to spontaneous symmetry breaking generating
(pseudo-)Goldstone bosons. A
spontaneously broken symmetry relates processes with different numbers
of Goldstone bosons. Therefore, the distinction between renormalizable 
($d\leq 4$) and nonrenormalizable ($d>4$) parts in the effective
Lagrangian (\ref{L_EFT}) becomes meaningless. The effective 
Lagrangian of type 3 is generically nonrenormalizable. Nevertheless, 
such Lagrangians define perfectly consistent quantum field theories
at sufficiently low energies. Instead
of the operator dimension as in (\ref{L_EFT}), the number of
derivatives of the fields and the number of symmetry breaking
insertions distinguish successive terms in the Lagrangian.
The general structure of effective Lagrangians with spontaneously broken
symmetries is largely independent of the specific physical
realization (universality). There are many examples in condensed
matter physics but the two main applications in particle physics are
electroweak symmetry breaking (Sec.~\ref{EWSB}) and chiral 
perturbation theory (Secs.~\ref{CHPT},\ref{NP}) with the spontaneously
broken global chiral symmetry of QCD.
\end{enumerate}  

Another classification of EFTs is related to the status of their
coupling constants.
\begin{enumerate} 
\item[A.] Coupling constants can be determined by matching the EFT
with the underlying theory at short distances. \\[.1cm]
The underlying theory is known and Green functions can be calculated
perturbatively at energies $\sim \La$ both in the fundamental and in
the effective theory. Identifying a minimal set of Green functions
fixes the couplings constants $g_{i_d}$ in Eq.~(\ref{L_EFT}) at the
scale $\La$. Renormalization group
equations can then be used to run the couplings down to lower
scales. The nonrenormalizable terms in the Lagrangian
(\ref{L_EFT}) can be fully included in the perturbative analysis.
\item[B.] Coupling constants are constrained by
symmetries only.
\begin{itemize} 
\item The underlying theory and therefore also the EFT coupling
constants are unknown. This is the case of the SM (Sec.~\ref{SMEFT}). 
A perturbative analysis beyond leading order only makes sense for the 
known renormalizable part $\cL_{d\leq 4}$. The nonrenormalizable terms 
suppressed by powers of $\La$ are
considered at tree level only. The associated coupling constants
$g_{i_d}$ serve as bookmarks for new physics. Usually, but
not always (cf., e.g., Sec.~\ref{NCQFT}), the symmetries of $\cL_{d\leq
4}$ are assumed to constrain the couplings.
\item The matching cannot be performed in perturbation theory even
  though the underlying theory is known. This is the generic situation
  for EFTs of type 3 involving spontaneous symmetry breaking. The
  prime example is chiral perturbation theory as the EFT of QCD at low       
  energies. 
\end{itemize}   
\end{enumerate}

\section{The Standard Model as an EFT}
\label{SMEFT}
With the possible exception of the scalar sector to be discussed in
Sec.~\ref{EWSB}, the SM is very likely the renormalizable part of an
EFT of type 1\,B. Except for nonzero neutrino masses, the SM
Lagrangian $\cL_{d\le 4}$ in (\ref{L_EFT}) accounts for physics up to
energies of roughly the Fermi scale $G_F^{-1/2} \simeq 300$ GeV.

Since the SM works exceedingly well up to the Fermi scale where the 
electroweak gauge symmetry is spontaneously broken it is natural to 
assume that the operators $O_{i_d}$ with $d>4$, made up from 
fields representing the known degrees of freedom and including a
single Higgs doublet in the SM proper, should be gauge invariant with 
respect to the full SM gauge group  
$SU(3)_c \times SU(2)_L \times U(1)_Y$. An almost obvious constraint
is Lorentz invariance that will be lifted in
Sec.~\ref{NCQFT}, however. 

These requirements limit the
Lagrangian with operator dimension d=5 to a single term (except for
generation multiplicity), consisting only of a left-handed lepton
doublet $L_L$ and the Higgs doublet $\Phi$:
\begin{equation} 
O_{d=5} = \eps_{ij} \eps_{kl}  L_{iL}^\top C^{-1} L_{kL}  
\Phi_j \,\Phi_l + ~{\rm h.c.}  
\end{equation}
This term violates lepton number and generates nonzero
Majorana neutrino masses. For a neutrino mass of 1 eV, the scale $\La$ 
would have to be of the order of $10^{13}$ GeV if the associated 
coupling constant in the EFT Lagrangian (\ref{L_EFT}) is of order 1. 

In contrast to the simplicity for d=5, the list of gauge invariant
operators with d=6 \,is enormous. Among them are operators violating
baryon or lepton number that must be associated with a scale much
larger than 1 TeV. To 
explore the territory close to present energies, it therefore makes
sense to impose baryon and lepton number conservation on the operators
with d=6. Those operators have all been classified (Buchm\"uller and
Wyler, 1986) and the number of
independent terms is of the order of 80. They can be grouped in three
classes. 

The first class consists of gauge and Higgs fields only. The
corresponding EFT Lagrangian has been used to parametrize new physics
in the gauge sector constrained by precision data from LEP.
The second class consists of operators bilinear in
fermion fields, with additional gauge and Higgs fields to generate
d=6. Finally, there are 4-fermion operators without 
other fields or derivatives. Some of the operators in the
last two groups are also constrained by precision experiments, with a
certain hierarchy of limits. For lepton and/or quark flavour
conserving terms, the best limits on $\La$ are in the few TeV range
whereas the absence of neutral flavour changing processes yields
lower bounds on $\La$ that are several orders of magnitude
larger. If there is new physics in the TeV range flavour changing
neutral transitions must be strongly suppressed, a powerful constraint 
on model building.

It is amazing that the most general renormalizable Lagrangian with the
given particle content accounts for almost all experimental results
in such an impressive manner. Finally, we recall that many of the
operators of dimension 6 are also generated in the SM via radiative
corrections. A necessary condition for detecting evidence for new
physics is therefore that the theoretical accuracy of radiative
corrections matches or surpasses the experimental precision.

\subsection{Noncommutative space-time}
\label{NCQFT}
Noncommutative geometry arises in some string theories and may be
expected on general grounds when incorporating gravity into a quantum 
field theory framework. The natural scale of
noncommutative geometry would be the Planck scale in this case without
observable consequences at presently accessible energies. However,
as in theories with large extra dimensions the characteristic
scale $\La_{\rm NC}$ could be significantly smaller. In parallel to
theoretical developments to define consistent noncommutative quantum
field theories (short for quantum field theories on noncommutative 
space-time), a number of phenomenological investigations have 
been performed to put lower bounds on $\La_{\rm NC}$. 

Noncommutative geometry is a deformation of ordinary space-time where
the coordinates, represented by hermitian operators $\hat{x}_\mu$, do
not commute:
\begin{equation} 
\left[\hat{x}_\mu,\hat{x}_\nu \right]=i\, \theta_{\mu\nu}~.
\end{equation}
The antisymmetric real tensor $\theta_{\mu\nu}$ has dimensions 
length$^2$ and it can be interpreted as parametrizing the resolution
with which space-time can be probed. In practically all applications,
$\theta_{\mu\nu}$ has been assumed to be a constant tensor and we may
associate an energy scale $\La_{\rm NC}$ with its non-zero entries:
\begin{equation} 
\La_{\rm NC}^{-2} \sim  \theta_{\mu\nu}~.
\end{equation}
 
There is to date no unique form for the noncommutative extension of
the SM. Nevertheless, possible observable effects of noncommutative
geometry have been investigated. Not unexpected
from an EFT point of view, for energies $\ll \La_{\rm NC}$ 
noncommutative field theories 
are equivalent to ordinary quantum field theories in the presence of
non-standard terms containing  $\theta_{\mu\nu}$ (Seiberg-Witten
map). Practically all applications have concentrated on effects linear
in  $\theta_{\mu\nu}$.

Kinetic terms in the Lagrangian are in general unaffected by the
noncommutative structure. New effects arise therefore mainly from 
renormalizable d=4 interactions terms. For example, the
Yukawa coupling $g_Y \overline{\psi} \psi \phi$ generates the
following interaction linear in $\theta_{\mu\nu}$:
\begin{equation} 
\cL_Y^{\rm NC} = g_Y \theta_{\mu\nu} \left(\partial^\mu
\overline{\psi} \partial^\nu \psi \phi + \partial^\mu
\overline{\psi}  \psi \partial^\nu \phi +  
\overline{\psi} \partial^\mu \psi \partial^\nu \phi\right)~.
\end{equation}
These interaction terms have operator dimension six and they are
suppressed by $\theta_{\mu\nu} \sim \La_{\rm NC}^{-2}$.
The major difference to the
previous discussion on physics beyond the SM is that there is an
intrinsic violation of Lorentz invariance due to the constant tensor
$\theta_{\mu\nu}$. In contrast to the previous analysis, the terms with
dimension d$\,>\,$4 do not respect the symmetries of the SM. 

If $\theta_{\mu\nu}$ is indeed constant over macroscopic distances 
many tests of Lorentz invariance can be used to put lower bounds on 
$\La_{\rm NC}$. Among the exotic effects investigated are
modified dispersion relations for particles, decay of high-energy
photons, charged particles producing Cerenkov radiation in vacuum,
birefringence of radiation, a variable speed of light, etc. A generic
signal of noncommutativity is the violation of angular momentum
conservation that can be searched for at LHC and at the next linear
collider.  

Lacking a unique noncommutative extension of the SM,
unambiguous lower bounds on $\La_{\rm NC}$ are difficult to
establish. However, the range $\La_{\rm NC} \lsim$ 10 TeV is almost
certainly excluded. An
estimate of the induced electric dipole moment of the electron
(noncommutative field theories
violate CP in general to first order in $\theta_{\mu\nu}$) yields 
$\La_{\rm NC} \gsim$ 100 TeV. On the other hand, if the SM were CP
invariant, noncommutative geometry
would be able to account for the observed CP violation in 
$K^0-\overline{K^0}$ mixing for $\La_{\rm NC} \sim$ 2 TeV.

\subsection{Electroweak symmetry breaking}
\label{EWSB}
In the SM, electroweak symmetry breaking is realized in the simplest
possible way through renormalizable interactions of a scalar Higgs
doublet with gauge bosons and fermions, a gauged version of the linear
$\sigma$ model. 

The EFT version of electroweak symmetry breaking (EWEFT) uses only the
experimentally established degrees of freedom in the SM (fermions and
gauge bosons). Spontaneous
gauge symmetry breaking is realized nonlinearly, without introducing 
additional scalar degrees of freedom. It
is a low-energy expansion where energies and masses are assumed to be 
small compared to the symmetry breaking scale. From both perturbative
and nonperturbative arguments we know that this scale cannot be much
bigger than 1 TeV.
The Higgs model can be viewed as a specific example of an EWEFT as 
long as the Higgs boson is not too light (heavy-Higgs scenario). 

The lowest-order effective Lagrangian takes the following form:
\begin{equation} 
\cL^{(2)}_{\rm EWSB} = \cL_B + \cL_F ~,
\label{L2_EWSB}
\end{equation}
where $\cL_F$ contains the gauge invariant kinetic terms for 
quarks and leptons including mass terms.
In addition to the kinetic terms for the gauge bosons $\vec W_\mu,
B_\mu$, the bosonic Lagrangian $\cL_B$ contains the characteristic
lowest-order term for the would-be-Goldstone bosons:
\begin{equation} 
\cL_B = \cL_{\rm gauge}^{\rm kin} + 
\frac{v^2}{4} \langle D_\mu U^\dg D^\mu U \rangle ~, 
\label{EWEFT_B} 
\end{equation}
with the gauge-covariant derivative 
\begin{eqnarray}   
D_\mu U &=& \partial_\mu U - i g W_\mu U + i g^\prime U
{\hat B}_\mu ~, \quad W_\mu = \frac{\vec \tau}{2} \vec W_\mu~, \quad 
{\hat B}_\mu = \frac{\tau_3}{2} B_\mu ~,
\end{eqnarray} 
where $\langle \dots \rangle$ denotes a (2-dimensional) trace.
The matrix field $U(\phi)$ carries the nonlinear
representation of the spontaneously broken gauge group and takes the
value $U=\mathbbm{1}$ in the unitary gauge.
The Lagrangian (\ref{L2_EWSB}) is invariant under local $SU(2)_L
\times U(1)_Y$ transformations:
\begin{eqnarray} 
W_\mu \lra g_L W_\mu g_L^\dg + \frac{i}{g} g_L \partial_\mu g_L^\dg
&,& {\hat B}_\mu \lra {\hat B}_\mu  + \frac{i}{g^\prime} g_R 
\partial_\mu g_R^\dg \\
f_L \lra g_L f_L ~, ~ f_R \lra g_R f_R &,& U \lra g_L U g_R^\dg~, \nn
\end{eqnarray} 
with $g_L(x)=\exp{(i \vec\alpha_L(x)
\vec\tau/2)}$, $g_R(x)=\exp{(i \alpha_Y(x) \tau_3/2)}$ and $f_{L(R)}$
are quark and lepton fields grouped in doublets.

As is manifest in the unitary gauge $U=\mathbbm{1}$, the lowest-order
Lagrangian of the EWEFT just implements the tree-level masses of gauge
bosons ($M_W = M_Z \cos{\theta_W} = v g/2, \tan{\theta_W}=g^\prime/g$) 
and fermions but does not carry any further information about
the underlying mechanism of spontaneous gauge symmetry breaking. This
information is first encoded in the couplings $a_i$ of the
next-to-leading-order Lagrangian
\begin{equation}  
\cL^{(4)}_{\rm EWSB} = \displaystyle\sum_{i=0}^{14} a_i O_i 
\label{L4_EWSB}
\end{equation}
with monomials $O_i$ of $O(p^4)$ in the low-energy expansion. The
Lagrangian (\ref{L4_EWSB}) is the most general CP and $SU(2)_L
\times U(1)_Y$ invariant Lagrangian of $O(p^4)$. 

Instead of listing the full Lagrangian, we display three typical
examples:
\begin{eqnarray}
O_0 = \frac{v^2}{4} \langle T V_\mu \rangle^2~, &
O_3 = - g \langle W_{\mu\nu} [V^\mu,V^\nu] \rangle ~, &
O_5 = \langle V_\mu V^\mu \rangle^2  ~,
\label{Oi_ex}
\end{eqnarray}  
where
\begin{eqnarray} 
T=U \tau_3 U^\dg ~, & V_\mu=D_\mu U U^\dg ~, & W_{\mu\nu}=
\frac{i}{g} \left[\partial_\mu - i g W_\mu, \partial_\nu - i g W_\nu
\right] ~. 
\end{eqnarray}
In the unitary gauge, the monomials $O_i$ reduce to polynomials in
the gauge fields. The three examples in Eq.~(\ref{Oi_ex}) start with
quadratic, cubic and quartic terms in the gauge fields, respectively. 
The strongest
constraints exist for the coefficients of quadratic contributions from
LEP1, less restrictive ones for the cubic self-couplings from LEP2 and
none so far for the quartic ones.

\section{Heavy quark physics}
\label{HQP}
EFTs in this section are derived from the SM and they are of type
2\,A in the classification of Sec.~\ref{class}. 
In a first step, one integrates out $W$, $Z$ and top quark.
Evolving down from $M_W$ to $m_b$, large logarithms 
$\alpha_s(m_b) \ln{(M_W^2/m_b^2)}$ are resummed into the Wilson
coefficients. At the scale of the
$b$-quark, QCD is still perturbative so that at least a part of the
amplitudes is calculable in perturbation theory. To separate the 
calculable part from the
rest, the EFTs below perform an expansion in $1/m_Q$ where
$m_Q$ is the mass of the heavy quark. 

Heavy quark EFTs offer several important advantages.
\begin{enumerate} 
\item[a.] Approximate symmetries that are hidden in full QCD appear in 
the expansion in $1/m_Q$.
\item[b.] Explicit calculations simplify in general, e.g., the
summing of large logarithms via renormalization group equations.
\item[c.] The systematic separation of hard and soft effects for
certain matrix elements (factorization) can be achieved much easier.
\end{enumerate} 

\subsection{Heavy quark effective theory}
\label{HQET}
Heavy quark effective theory (HQET) is reminiscent of 
the Foldy-Wouthuy\-sen transformation (nonrelativistic
expansion of the Dirac equation). It is a systematic expansion in
$1/m_Q$  when $m_Q \gg \La_{\rm QCD}$, the scale parameter of QCD. It 
can be applied to
processes where the heavy quark remains essentially on shell: its
velocity $v$ changes only by small amounts $\sim \La_{\rm
QCD}/m_Q$. In the hadron rest frame, the heavy quark is almost at rest
and acts as a quasi-static source of gluons.

More quantitatively, one writes the heavy quark momentum as
$p^\mu=m_Q\,v^\mu + k^\mu$ where $v$ is the hadron four-velocity
($v^2=1$) and $k$ is a residual momentum of $O(\La_{\rm QCD})$. The
heavy quark field $Q(x)$ is then decomposed with the help of energy 
projectors $P_v^\pm=(1 \pm \slch{v})/2$ and employing a field
redefinition:
\begin{eqnarray}
\label{redef}
Q(x) = e^{-i m_Q v\cdot x} \left(h_v(x) + H_v(x) \right) & &\\
h_v(x) = e^{i m_Q v\cdot x} P_v^+ Q(x)~, & & H_v(x) = 
e^{i m_Q v\cdot x} P_v^- Q(x)~. \nn
\end{eqnarray}
In the hadron rest frame, $h_v(x)$ and $H_v(x)$ correspond to the
upper and lower components of $Q(x)$, respectively. With this
redefinition, the heavy-quark Lagrangian is expressed in terms of 
a massless field $h_v$ and a ``heavy'' field $H_v$:
\begin{eqnarray} 
\cL_Q &=& \overline{Q} (i \slch{D} - m_Q) Q \nl
&=& \overline{h_v} \,i v \cdot D \,h_v - \overline{H_v} ( i v\cdot D + 2
m_Q ) H_v + {\rm mixed~~terms}~.
\end{eqnarray}
At the semi-classical level, the
field $H_v$ can be eliminated by using the QCD field equation $(i
\slch{D} - m_Q) Q=0$ yielding the nonlocal expression
\begin{equation} 
\cL_Q = \overline{h_v} \,i v \cdot D \,h_v + \overline{h_v} \,i
\slch{D}_\perp \frac{1}{i v \cdot D + 2 m_Q - i\eps} i \slch{D}_\perp
h_v  \label{L_Q}
\end{equation}
with $D^\mu_\perp=(g^{\mu\nu}-v^\mu v^\nu) D_\nu$. The field
redefinition in (\ref{redef}) ensures that in the
heavy-hadron rest frame derivatives of $h_v$ give rise to small
momenta of $O(\La_{\rm QCD})$ only.
The Lagrangian (\ref{L_Q}) is the starting point for a systematic
expansion in $m_Q$. 

To leading order in $1/m_Q ~(Q=b,c)$, the Lagrangian
\begin{equation}
\cL_{b,c}= \overline{b_v} \,i v \cdot D \,b_v + 
\overline{c_v} \,i v \cdot D \,c_v 
\label{HQETsymm} 
\end{equation}
exhibits two important approximate symmetries of HQET: the flavour
symmetry $SU(2)_F$ relating heavy quarks moving with the same velocity
and the heavy-quark spin symmetry generating an overall $SU(4)$
spin-flavour symmetry. The flavour symmetry is obvious and
the spin symmetry is due to the absence of Dirac matrices in
(\ref{HQETsymm}): both spin degrees of freedom couple to gluons in the
same way. The simplest spin-symmetry doublet consists of a pseudoscalar
meson $H$ and the associated vector meson $H^*$. Denoting the doublet
by $\cH$, the matrix elements of the heavy-to-heavy transition current 
are determined to leading order in $1/m_Q$ by a single form factor, 
up to Clebsch-Gordan coefficients:
\begin{equation} 
\langle \cH(v^\prime) | \overline{h_{v^\prime}} \Gamma h_v |\cH(v)
\rangle \sim \xi(v\cdot v^\prime)~.
\end{equation}
$\Gamma$ is an arbitrary combination of Dirac matrices and the form
factor $\xi$ is the so-called Isgur-Wise function. Moreover, since
$\overline{h_v} \gamma^\mu h_v$ is the Noether current of 
heavy flavour symmetry, the Isgur-Wise function is fixed in the
no-recoil limit $v^\prime=v$ to be $\xi(v\cdot v^\prime =1)=1$. 
The semileptonic decays $B \to D l \nu_l$ and $B \to D^\star l \nu_l$
are therefore governed by a single normalized form factor to leading
order in $1/m_Q$, with important consequences for the determination of
the CKM matrix element $V_{cb}$. 

The HQET Lagrangian is superficially frame dependent. Since the SM is
Lorentz invariant the HQET Lagrangian must be independent of
the choice of the frame vector $v$. Therefore, a shift in $v$
accompanied by corresponding shifts of the fields $h_v$ and of the
covariant derivatives must leave the Lagrangian invariant. This
reparametrization invariance is unaffected by renormalization and it
relates coefficients with different powers in $1/m_Q$.

\subsection{Soft collinear effective theory}
\label{SCET}
HQET is not applicable in heavy quark decays where some of the light 
particles in the final state have momenta of $O(m_Q)$, e.g., for
inclusive decays like $B \to X_s \gamma$ or exclusive ones like $B
\to \pi \pi$. In recent years, a systematic heavy quark expansion for 
heavy-to-light decays has been set up in the form of soft collinear 
effective theory (SCET). 

SCET is more complicated than HQET because now the low-energy theory
involves more than one scale. In the SCET Lagrangian a light
quark or gluon field is represented by several effective fields.
In addition to the soft fields $h_v$ in
(\ref{L_Q}), so-called collinear fields enter that have large energy
and carry large momentum in the direction of the light hadrons in the
final state. 

In addition to the frame vector $v$ of HQET ($v=(1,0,0,0)$ in the
heavy-hadron rest frame), SCET introduces a light-like reference
vector $n$ in the direction of the jet of energetic light particles
(for inclusive decays), e.g., $n=(1,0,0,1)$. All momenta $p$ are
decomposed in terms of light-cone coordinates $(p_+,p_-,p_\perp)$ with
\begin{equation}
p^\mu = \frac{n\cdot p}{2}\,\overline{n}^\mu +  \frac{\overline{n} 
\cdot p}{2}\,n^\mu + p^\mu_\perp= p_+^\mu + p_-^\mu + p_\perp^\mu
\end{equation}
where $\overline{n} = 2 v - n =(1,0,0,-1)$. For large energies the
three light-cone components are widely separated, with $p_-=O(m_Q)$
being large while $p_\perp$ and $p_+$ are small. Introducing a small
parameter $\lambda \sim p_\perp/p_-$, the light-cone components of
(hard-)collinear particles scale like
$(p_+,p_-,p_\perp)=m_Q(\lambda^2,1,\lambda)$. Thus, there are three
different scales in the problem compared to only two in HQET. 
For exclusive decays, the situation is even more involved.

The SCET Lagrangian is obtained from the full theory by an expansion
in powers of $\lambda$. In addition to the heavy quark field $h_v$,
one introduces soft as well as collinear quark
and gluon fields by field redefinitions so that the various
fields have momentum components that scale appropriately with
$\lambda$. 

Similar to HQET, the leading-order Lagrangian of SCET exhibits again
approximate symmetries that can lead to a reduction of form factors
describing heavy-to-light decays. As in HQET,
reparametrization invariance implements Lorentz invariance and results
in stringent constraints on subleading corrections in SCET.

An important result of SCET is the proof of factorization theorems to
all orders in $\alpha_s$. For inclusive decays, the differential rate
is of the form
\begin{equation}
d\Gamma \sim H J \times S
\end{equation}
where $H$ contains the hard corrections. The so-called jet function
$J$ sensitive to the collinear region is convoluted with the shape
function $S$ representing the soft contributions. At leading order,
the shape function drops out in the ratio of weighted decay spectra
for $B \to X_u l \nu_l$ and $B \to X_s \gamma$ allowing for a
determination of the CKM matrix element $V_{ub}$. Factorization
theorems have become available for an increasing number of processes,
most recently also for exclusive decays of $B$ into two light
mesons.

\subsection{Nonrelativistic QCD}
\label{NRQCD}

In HQET the kinetic energy of the heavy quark appears as a small
correction of $O(\Lambda^2_{\rm QCD}/m_Q)$. For systems with more than 
one heavy quark the kinetic energy cannot be treated as a perturbation
in general. For instance, the virial theorem implies that 
the kinetic energy in quarkonia $\overline{Q} Q$ is of the same order 
as the binding energy of the bound state.

NRQCD, the EFT for heavy quarkonia, is an extension of HQET. The 
Lagrangian for NRQCD coincides with HQET in the bilinear sector of the 
heavy quark fields but it includes also quartic interactions between
quarks and antiquarks. The relevant expansion parameter in this case
is the relative velocity between $Q$ and $\overline{Q}$. In contrast
to HQET, there are 
at least three widely separate scales in heavy quarkonia: in addition
to $m_Q$, the relative momentum of the bound quarks $p \sim m_Q v$
with $v \ll 1$ and the typical kinetic energy $E \sim m_Q
v^2$. The main challenges are to derive the quark-antiquark
potential directly from QCD and to describe quarkonium production and
decay at collider experiments.  In the Abelian case, the corresponding
EFT for QED is called NRQED that has been used to study
electromagnetically bound systems like the hydrogen atom, positronium,
muonium, etc.

In NRQCD only the hard degrees of freedom with momenta $\sim m_Q$ are
integrated out. Therefore, NRQCD is not enough for a systematic
computation of heavy quarkonium properties. Because the
nonrelativistic fluctuations of order $m_Q v$ and $m_Q v^2$ have not
been separated, the power counting in NRQCD is ambiguous in higher
orders. 

To overcome those deficiencies, two approaches have been put forward: 
p(o\-ten\-tial)NRQCD and v(elocity)NRQCD. In pNRQCD, a two-step procedure
is employed for integrating out quark and gluon degrees of freedom:
\begin{eqnarray} 
{\rm QCD} & \mbox{ } \hspace*{2cm} & \mbox{ } \hspace*{.3cm}\La > m_Q
\nl 
&  \Downarrow & \nl
{\rm NRQCD} & & m_Q > \La > m_Q v \nl
&  \Downarrow & \nl
{\rm pNRQCD} & & m_Q v > \La > m_Q v^2 \nn
\end{eqnarray}
The resulting EFT derives its name from the fact that the four-quark
interactions generated in the matching procedure are the potentials
that can be used in Schr\"odinger perturbation theory. It is claimed
that pNRQCD can also be used in the nonperturbative domain where
$\alpha_s(m_Q v^2)$ is of order one or larger. The advantage would be
that also charmonium becomes accessible to a systematic EFT analysis. 

The alternative approach of vNRQCD is only applicable in the fully
perturbative regime when $m_Q \gg
m_Q v \gg m_Q v^2 \gg \La_{\rm QCD}$ is valid. It separates the
different degrees of freedom in a single step leaving only ultrasoft 
energies and momenta of $O(m_Q v^2)$ as continuous variables. The 
separation of larger scales proceeds in a similar 
fashion as in HQET via field redefinitions. A systematic
nonrelativistic power counting in the velocity $v$ is implemented.

\section{The Standard Model at low energies}
\label{SMLE}
At energies below 1 GeV hadrons rather than quarks and gluons are the
relevant degrees of freedom. Although the strong interactions are highly
nonperturbative in the confinement region Green functions and amplitudes
are amenable to a systematic low-energy expansion. The key observation
is that the QCD Lagrangian with $N_f=$ 2 or 3 light quarks,
\begin{eqnarray} 
\cL_{\rm QCD} &=& \overline{q} \left( i \Dslash - \cM_q \right) q
- {1\over 4}G^\alpha_{\mu\nu} G^{\alpha\mu\nu} + 
{\cal L}_{\mbox{\tiny heavy quarks}} \label{QCD}\\*
&=& \overline{q_L} i \Dslash q_L + \overline{q_R} i \Dslash q_R 
- \overline{q_L} \cM_q q_R - \overline{q_R} \cM_q q_L
+ \dots ~,\nl
q_{R,L} &=& {1\over 2}(1 \pm \gamma_5)q ~,\hspace*{2cm}
q^\top=( u \, d \, [s]) ~,\nn 
\end{eqnarray} 
exhibits a global symmetry 
\begin{equation}
\underbrace{SU(N_f)_L \times SU(N_f)_R}_{\mbox{chiral group $G$}}
\times U(1)_V \times U(1)_A  
\end{equation}
in the limit of $N_f$ massless quarks ($\cM_q=0$). 
At the hadronic level,  the quark number symmetry $U(1)_V$ is realized 
as baryon number. The axial $U(1)_A$ is not a symmetry at the quantum 
level due to the  Abelian anomaly. 

Although not yet derived from first principles, there are 
compelling theoretical and phenomenological arguments that the ground
state of QCD is not even approximately chirally symmetric. All
evidence, such as the existence of relatively light pseudoscalar
mesons, points to spontaneous chiral symmetry breaking $G \lra
SU(N_f)_V$ where $SU(N_f)_V$ is the diagonal subgroup of $G$. The
resulting $N_f^2 - 1$ (pseudo-)Goldstone bosons interact weakly at low
energies. In fact, Goldstone's theorem ensures that purely mesonic or
single-baryon amplitudes vanish in the chiral limit ($\cM_q=0$) when
the momenta of all pseudoscalar mesons tend to zero. This is the basis
for a systematic low-energy expansion of Green functions and
amplitudes. The corresponding EFT (type 3\,B in
the classification of Sec.~\ref{class}) is called chiral perturbation 
theory (CHPT) (Weinberg, 1979; Gasser and Leutwyler, 1984, 1985).

Although the construction of effective Lagrangians with nonlinearly 
realized chiral symmetry is well understood there are some subtleties 
involved. First of all, there may be terms in a chiral 
invariant action that cannot be written as the four-dimensional
integral of an invariant Lagrangian. The chiral anomaly for
$SU(3)\times SU(3)$ bears witness of this fact and gives rise to the
Wess-Zumino-Witten action. A general theorem to account for such
exceptional cases is due to D'Hoker and Weinberg
(1994). Consider the most general action for Goldstone fields with
symmetry group $G$, spontaneously broken to a subgroup $H$. The only
possible non-$G$-invariant terms in the Lagrangian that give rise to
a $G$-invariant action are in one-to-one correspondence with the
generators of the fifth cohomology group $\cH^5(G/H;\mathbbm{R})$ of
the coset manifold $G/H$. For the relevant case of chiral $SU(N)$, the 
coset space $SU(N)_L\times SU(N)_R/SU(N)_V$ is itself an $SU(N)$
manifold. For $N \ge 3$, $\cH^5(SU(N);\mathbbm{R})$ has a single
generator that corresponds precisely to the Wess-Zumino-Witten term. 

At a still deeper level, one may ask whether chiral
invariant Lagrangians are sufficient (except for the anomaly) to 
describe the low-energy structure of Green functions as dictated by 
the chiral Ward identities of QCD. To be able to calculate such Green
functions in general, the global chiral symmetry of QCD is
extended to a local symmetry by the introduction of external gauge 
fields. The following invariance theorem (Leutwyler, 1994) provides an 
answer to the above question. Except for the anomaly, the
most general solution of the Ward identities for a spontaneously
broken symmetry in Lorentz invariant theories can be obtained from
gauge invariant Lagrangians to all orders in the low-energy
expansion. The restriction to Lorentz invariance is crucial:
the theorem does not hold in general in nonrelativistic effective
theories.   

\subsection{Chiral perturbation theory}
\label{CHPT}
The effective chiral Lagrangian of the SM in the meson sector is
displayed in Table \ref{EFTSM}. The lowest-order Lagrangian for the
purely strong interactions is given by
\begin{equation} 
\cL_{p^2}  = \frac{F^2}{4} \langle D_\mu U D^\mu U^\dagger \rangle + 
\displaystyle\frac{F^2 B}{2} \langle (s+ip) U^\dagger + 
(s-ip)  U  \rangle ~,
\label{L_2}
\end{equation} 
with a covariant derivative $D_\mu U = \partial_\mu U - 
i (v_\mu + a_\mu) U + i U (v_\mu - a_\mu)$. The first term has the
familiar form (\ref{EWEFT_B}) of the gauged nonlinear $\sigma$ model, 
with the matrix field $U(\phi)$ transforming as $U \lra g_R U g_L^\dg$
under chiral rotations. External fields $v_\mu, a_\mu, s, p$ are
introduced for constructing the generating functional of Green
functions of 
quark currents. To implement explicit chiral symmetry breaking, the
scalar field $s$ is set equal to the quark mass matrix $\cM_q$ at the 
end of the calculation. 

\renewcommand{\arraystretch}{1.2}
\begin{table}[ht]
\begin{center}
\caption{The effective chiral Lagrangian of the SM in the
  meson sector. The numbers in brackets refer to the number of
independent couplings for $N_f=3$. 
The parameter-free Wess-Zumino-Witten action $S_{\rm WZW}$
  that cannot be written as the four-dimensional integral of an
  invariant Lagrangian must be added.}
\label{EFTSM}
\vspace{.5cm}
\begin{tabular}{|l|c|} 
\hline
&  \\
\hspace{1cm} ${\cal L}_{\rm chiral\; order}$ 
~($\#$ of LECs)  &  loop  ~order \\[8pt] 
\hline 
&  \\
${\cal L}_{p^2}(2)$
~+~${\cal L}_{G_Fp^2}^{\Delta S=1}(2)$  
~+~${\cal L}_{e^2p^0}^{\rm em}(1)$
~+~${\cal L}_{G_8e^2p^0}^{\rm emweak}(1)$ & $L=0$ \\[15pt]
~+~${\cal L}_{p^4}(10)$~+~${\cal L}_{p^6}^{\rm odd}(32)$
~+~${\cal L}_{G_8p^4}^{\Delta S=1}(22)$
~+~${\cal L}_{G_{27}p^4}^{\Delta S=1}(28)$ &   
$L=1$ \\[3pt]
~+~${\cal L}_{e^2p^2}^{\rm em}(14)$
~+~${\cal L}_{G_8e^2p^2}^{\rm emweak}(14)$ 
~+~${\cal L}_{e^2p}^{\rm leptons}(5)$  & \\[15pt] 
~+~${\cal L}_{p^6}(90)$  & $L=2$ \\[8pt] 
\hline
\end{tabular}
\end{center}
\end{table}

The
leading-order Lagrangian has two free parameters $F,B$ related to the
pion decay constant and to the quark condensate, respectively:
\begin{eqnarray} 
F_\pi &=& F \left[1+O(m_q)\right]  \\
\langle 0| \overline{u}u|0\rangle &=& - F^2  B
\left[1+O(m_q)\right]~. \nn
\end{eqnarray}
The Lagrangian (\ref{L_2}) gives rise to $M_\pi^2= B(m_u+m_d)$ at
lowest order. From detailed studies of pion-pion scattering
(Colangelo, Gasser and Leutwyler, 2001) we know that the leading term
accounts for at least 94 \% of the pion mass. This supports the
standard counting of CHPT, with quark masses 
booked as $O(p^2)$ like the two-derivative term in (\ref{L_2}).

The effective chiral Lagrangian in Table \ref{EFTSM} contains the
following parts:
\begin{enumerate}  
\item[i.] Strong interactions: \quad  ${\cal L}_{p^2}$,  ${\cal L}_{p^4}$, 
${\cal L}_{p^6}^{\rm odd}$, ${\cal L}_{p^6} + S_{\rm WZW}$  
\item[ii.] Nonleptonic weak interactions to first order
  in the Fermi coupling constant $G_F$: \quad
${\cal L}_{G_Fp^2}^{\Delta S=1}$, ${\cal L}_{G_8 p^4}^{\Delta
  S=1}$, ${\cal L}_{G_{27}p^4}^{\Delta S=1}$ 
\item[iii.] Radiative corrections for strong processes: \quad
  ${\cal L}_{e^2p^0}^{\rm em}$, ${\cal L}_{e^2p^2}^{\rm em}$ 
\item[iv.] Radiative corrections for nonleptonic weak decays: 
\quad ${\cal L}_{G_8e^2p^0}^{\rm emweak}$, 
${\cal L}_{G_8e^2p^2}^{\rm emweak}$ 
\item[v.] Radiative corrections for semileptonic weak decays:
 \quad ${\cal L}_{e^2p}^{\rm leptons}$
\end{enumerate} 

\begin{figure}[t]
\centerline{\epsfig{file=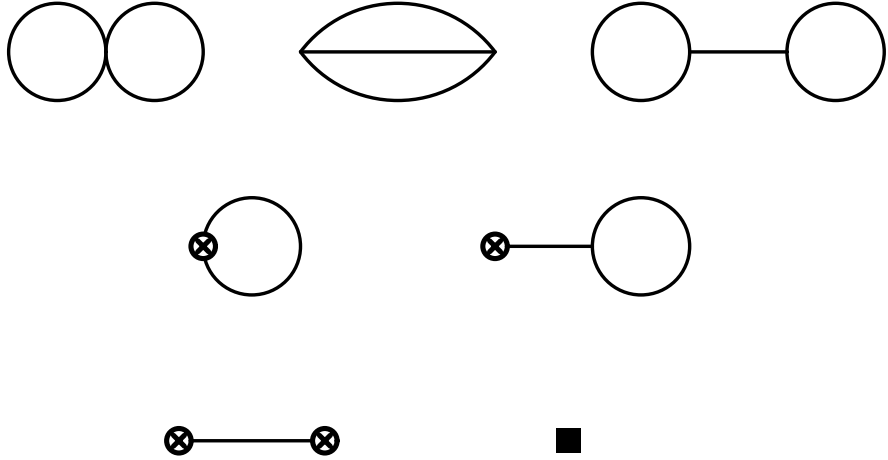,height=6cm}}
\caption{Skeleton diagrams of $O(p^6)$. Normal vertices are from
$\cL_{p^2}$, crossed circles and the full square denote vertices from 
$\cL_{p^4}$ and $\cL_{p^6}$, respectively.}
\label{p6diag}
\end{figure}

Beyond leading order, unitarity and analyticity require the inclusion
of loop contributions. In the purely strong sector, calculations have
been performed up to NNLO.
Fig.~\ref{p6diag} shows the corresponding skeleton
diagrams of $O(p^6)$, with full lowest-order tree structures to be
attached to propagators and vertices.  
The coupling
constants of the various Lagrangians in Table \ref{EFTSM} absorb the
divergences from loop diagrams leading to finite renormalized Green
functions with scale dependent couplings, the so-called low-energy
constants (LECs). As in all EFTs, the LECs parametrize the effect of
``heavy'' degrees of freedom that are not represented explicitly in
the EFT Lagrangian.
Determination of those LECs is a major task for
CHPT. In addition to phenomenological information, further theoretical
input is needed. Lattice gauge theory has already furnished values for
some LECs.
To bridge the gap between the low-energy domain of
CHPT and the perturbative domain of QCD, large-$N_c$ motivated
interpolations with meson resonance exchange have been used
successfully to pin down some of the LECs. 

Especially in cases where the knowledge of LECs is
limited, renormalization group methods provide valuable
information. As in renormalizable quantum field theories, the leading
chiral logs $(\ln{M^2/\mu^2})^L$ with a typical meson mass 
$M$, renormalization scale $\mu$ and loop order $L$ can in
principle be determined from one-loop diagrams only (B\"uchler and 
Colangelo, 2003). In contrast to
the renormalizable situation, new derivative structures (and quark
mass insertions) occur at each loop order preventing a straightforward
resummation of chiral logs.

Among the many applications of CHPT in the meson sector are the
determination of quark mass ratios and the analysis of pion-pion 
scattering where the chiral amplitude of NNLO has been combined with 
dispersion theory (Roy equations). Of increasing importance for
precision physics (CKM matrix elements, $(g-2)_\mu$, \dots) are
isospin violating corrections including radiative corrections where
CHPT provides the only reliable approach in the low-energy
region. Such corrections are also essential for the analysis of
hadronic atoms like pionium, a $\pi^+\pi^-$ bound state.

CHPT has also been applied extensively in the single-baryon sector. 
There are several differences to the purely mesonic case. For
instance, the 
chiral expansion proceeds more slowly and the nucleon mass $m_N$ 
provides a new scale that does not vanish in the chiral limit. The 
formulation of heavy baryon CHPT was modeled after HQET integrating 
out the nucleon modes of $O(m_N)$. To improve the convergence of the 
chiral expansion in some regions of phase space,
a manifestly Lorentz invariant formulation has been set up
more recently (relativistic baryon CHPT). Many single-baryon processes
have been calculated to N$^3$LO in both approaches, e.g., pion-nucleon 
scattering. With similar methods as in the mesonic sector,
hadronic atoms like pionic or kaonic hydrogen have been investigated.

\subsection{Nuclear physics}
\label{NP}
In contrast to the meson and single-baryon sectors, amplitudes with
two or more nucleons do not vanish in the chiral limit when the
momenta of Goldstone mesons tend to zero. Consequently, the power
counting is different in the many-nucleon sector. 
Multi-nucleon processes are treated with different EFTs
depending on whether all momenta are smaller or larger than the pion
mass. 

In the very-low-energy regime $|\vec p| \ll M_\pi$, pions or other
mesons do not appear as dynamical degrees of freedom. 
The resulting EFT is called ``pionless EFT''
and it describes systems like the deuteron where the typical nucleon
momenta are $\sim \sqrt{m_N B_d} \simeq 45$ MeV ($B_d$ is the
binding energy of the deuteron). The Lagrangian for the strong
interactions between two nucleons has the form
\begin{equation} 
\cL_{NN} = C_0 \left( N^\top P_i N \right)^\dg N^\top P_i N + \dots
\end{equation} 
where $P_i$ are spin-isospin projectors and higher-order terms contain 
derivatives of the nucleon fields. The existence of bound states
implies that at least part of the EFT Lagrangian must be treated
nonperturbatively. Pion\-less EFT is an extension of
effective range theory that has long been used in nuclear physics. It
has been applied successfully especially to the 
deuteron but also to more complicated few-nucleon systems like the
$Nd$ and $n\alpha$ systems. For instance, precise results for $Nd$
scattering have been obtained with parameters fully determined from
$NN$ scattering. Pionless EFT has also been applied to so-called halo
nuclei where a tight cluster of nucleons (like $^4$He) is surrounded
by one or more ``halo'' nucleons. 

In the regime $|\vec p| > M_\pi$, the pion must be included as a
dynamical degree of freedom. With some modifications in the power 
counting, the corresponding EFT is based on the
approach of Weinberg (1990,1991) who applied the usual rules of the meson 
and single-nucleon sectors to the nucleon-nucleon potential (instead
of the scattering amplitude). The potential is then to be inserted
into a Schr\"odinger equation to calculate physical observables. The
systematic power counting leads to a natural hierarchy of nuclear
forces, with only two-nucleon forces appearing up to NLO. Three- and
four-nucleon forces arise at NNLO and N$^3$LO, respectively.

A lot of progress has been achieved in the phenomenology of few-nucleon
systems. The two- and $n$-nucleon ($3 \le
n \le 6$) sectors have been pushed to N$^3$LO and NNLO, respectively,
with encouraging signs of ``convergence''. Compton scattering off 
the deuteron, $\pi d$ scattering, nuclear parity violation, solar fusion
and other processes have been investigated in the EFT
approach. The quark mass dependence of the nucleon-nucleon interaction
has also been studied.

\section*{Acknowledgements}
I thank N. Brambilla, W. Grimus, H. Grosse, R. Kaiser and A. Vairo for
helpful comments. This work was supported in part by
HPRN-CT2002-00311 (EURIDICE).

\section*{Further reading}
Appelquist, T. and Carazzone, J. (1975), Infrared singularities and
massive fields, Phys. Rev. D11, 2856.\\[.15cm]
Beane, S.R. et al. (2000), From hadrons to nuclei: crossing the border,
in  Boris Ioffe Festschrift ``At the frontier of particle
physics'', Ed. M. Shifman and B. Ioffe, World 
Scientific (Singapore, 2001) [nucl-th/0008064].\\[.15cm]
Becher, T. (2004), B decays in the heavy-quark expansion,
hep-ph/0411065.\\[.15cm]  
Bedaque, P.F. and van Kolck, U. (2002), Effective field theory for 
few-nucleon systems, Ann. Rev. Nucl. Part. Sci. 52, 339
[nucl-th/0203055].\\[.15cm]
Bernard, V., Kaiser, N. and Mei\ss ner, U.-G. (1995), Chiral dynamics in
nucleons and nuclei, Int. J. Mod. Phys. E4, 193
[hep-ph/9501384]. \\[.15cm] 
Bijnens, J. (2002), QCD and weak interactions of light quarks, in 
``At the frontier of particle physics'', Vol. 4,
Ed. M. Shifman, World Scientific (Singapore, 2002)
[hep-ph/0204068].\\[.15cm]
Bijnens, J. (2004), Chiral meson physics at two loops, 
hep-ph/0409068.\\[.15cm] 
Brambilla, N., Pineda, A., Soto, J. and Vairo, A. (2004), Effective
field theories for quarkonium, Rev. Mod. Phys. (in press), 
hep-ph/0410047.\\[.15cm]
Brambilla, N. et al. (Quarkonium Working Group) (2004), Heavy
quarkonium physics, hep-ph/0412158. \\[.15cm] 
B\"uchler, M. and Colangelo, G. (2003), Renormalization group
equations for effective field theories, Eur. Phys. J. C32, 427
[hep-ph/0309049].\\[.15cm]
Buchm\"uller, W. and Wyler, D. (1986), Effective Lagrangian analysis
of new interactions and flavour conservation, Nucl. Phys. B268,
621.\\[.15cm]
Colangelo, G., Gasser, J. and Leutwyler, H. (2001), $\pi\pi$
scattering, Nucl. Phys. B603, 125 [hep-ph/0103088].\\[.15cm]
D'Hoker, E. and Weinberg, S. (1994), General effective actions,
Phys. Rev. D50, 6050 [hep-ph/9409402].\\[.15cm]
Ecker, G. (1995), Chiral perturbation theory, 
Prog. Part. Nucl. Phys. 35, 1 [hep-ph/9501357].\\[.15cm]
Ecker, G. (1998), Chiral symmetry, in
  Proc. of Schladming Winter School 1998: ``Broken symmetries'',
  Eds. L. Mathelitsch and W. Plessas, Lecture Notes in Physics 521, 
  Springer (Berlin, 1999) [hep-ph/9805500].\\[.15cm]
Ecker, G. (2000), Strong interactions of light flavours, in Proc. of
Advanced School on Quantum Chromodynamics, Benasque, Spain, Eds. S. 
Peris and V. Vento, Univ. Autonoma de Barcelona (Barcelona, 2001) 
[hep-ph/0011026]. \\[.15cm] 
Feruglio, F. (1993), The chiral approach to the electroweak
interactions, Int. J. Mod. Phys. A8, 4937 [hep-ph/9301281].\\[.15cm]
Fleming, S. (2003), The large energy expansion for B decays: soft
collinear effective theory, in Proc. of 
``Flavour Physics and CP Violation'', Paris, Ed. P. Perret,
Ecole Polytechnique (Palaiseau, 2003) [hep-ph/0309133]. \\[.15cm]  
Gasser, J. and Leutwyler, H. (1984), Chiral perturbation theory
to one loop, Ann. Phys. 158, 142. \\[.15cm] 
Gasser, J. and Leutwyler, H. (1985), Chiral perturbation theory: 
expansions in the mass of the strange quark, Nucl. Phys. B250,
465.\\[.15cm]   
Gasser, J. (2003), Light quark dynamics, 
in Proc. of Schladming Winter School 2003: ``Flavour physics'',
  Eds. U.-G. Mei\ss ner and W. Plessas, Lecture Notes in Physics 629, 
  Springer (Berlin, 2004) [hep-ph/0312367].\\[.15cm]
Georgi, H. (1993), Effective field theory,
Ann. Rev. Nucl. Part. Sci. 43, 209. \\[.15cm] 
Hill, C. T. and Simmons, E. H. (2003), Strong dynamics and
electroweak symmetry breaking, Phys. Rept. 381, 235, Err. ibid. 390,
553 (2004) [hep-ph/0203079]. \\[.15cm] 
Hinchliffe, I., Kersting, N. and Ma, Y.L. (2004), Review of the
phenomenology of noncommutative geometry, Int. J. Mod. Phys. A19, 179
[hep/0205040]. \\[.15cm]
Hoang, A.H. (2002), Heavy quarkonium dynamics, in ``At the frontier of 
particle physics'', Vol. 4,  Ed. M. Shifman, World 
Scientific (Singapore, 2002) [hep-ph/0204299].\\[.15cm]
Kaplan, D.B. (1995), Effective field theories,
  Lectures at 7th Summer School in Nuclear Physics, Seattle,
  Washington, nucl-th/9506035.\\[.15cm]
Leutwyler, H. (1994), On the foundations of chiral
perturbation theory, Ann. Phys. 235, 165 [hep-ph/9311274].\\[.15cm]
Leutwyler, H. (2000), Chiral dynamics,
in  Boris Ioffe Festschrift: ``At the frontier of particle
physics'', Ed. M. Shifman and B. Ioffe, World 
Scientific (Singapore, 2001) [hep-ph/0008124]. \\[.15cm] 
Mannel, T. (2004), Effective field theories in flavour physics,
Springer Tracts in Modern Physics 203, Springer (Berlin, 2004).  \\[.15cm]
Manohar, A.V. (1996), Effective field theories, in
  Proc. of Schladming Winter School 1996: ``Perturbative and
  nonperturbative aspects of quantum field theory'', Eds. H. Latal and
  W. Schweiger, Lecture Notes in Physics 479, Springer (Berlin, 1997) 
  [hep-ph/9606222].\\[.15cm]
Manohar, A.V. and Wise, M.B. (2000), Heavy quark physics,
Camb. Monogr. Part. Phys. Nucl. Phys. Cosmol. 10, 1.\\[.15cm]
Mei\ss ner, U.-G. (2005), Modern theory of nuclear forces,
Nucl. Phys. A751, 149 [nucl-th/0409028]. \\[.15cm]
Neubert, M. (2004), Theory of exclusive hadronic B decays,  
Lecture Notes in Physics 647, Springer (Berlin, 2004). \\[.15cm] 
Pich, A. (1995), Chiral perturbation theory, Rept. Prog. Phys. 58, 563 
[hep-ph/9502366]. \\[.15cm] 
Pich, A. (1998), Effective field theory, in Proc. of
Les Houches Summer School 1997: ``Probing the standard model of 
particle interactions'', Eds. R. Gupta et al., Elsevier (Amsterdam,
1999) [hep-ph/9806303].\\[.15cm]
Rafael, E. de (1995), Chiral Lagrangians and CP violation, 
in Proc. of TASI 1994: ``CP violation and the limits of the Standard
Model'', Ed. J. F. Donoghue, World Scientific (River Edge, 1995)
[hep-ph/9502254]. \\[.15cm] 
Rothstein, I.Z. (2003), Effective field theories, in
  Proc. of TASI 2002: ``Particle physics and cosmology: the quest for 
 physics beyond the standard model'', Eds. H.E. Haber and A.E. Nelson, 
World Scientific (River Edge, 2004) [hep-ph/0308266].\\[.15cm]
Savage, M. (2003), Effective field theory for nuclear physics,
nucl-th/0301058. \\[.15cm] 
Scherer, S. (2002), Introduction to chiral perturbation theory, 
Adv. Nucl. Phys. 27, 277, Ed. J.W. Negele et
al. [hep-ph/0210398].\\[.15cm]
Stewart, I.W. (2003), Theoretical introduction to B decays and the
soft collinear effective theory, hep-ph/0308185. \\[.15cm] 
Szabo, R.J. (2003), Quantum field theory in noncommutative spaces,
Phys. Rept. 378, 207  [hep-th/0109162].\\[.15cm]
Weinberg, S. (1979), Phenomenological Lagrangians, 
Physica A96, 327.\\[.15cm]
Weinberg, S. (1990), Nuclear forces from chiral Lagrangians,
Phys. Lett. B251, 288. \\[.15cm] 
Weinberg, S. (1991), Effective chiral Lagrangians for
nucleon-pion interactions and nuclear forces, 
Nucl. Phys. B363, 3.

\end{document}